\documentclass [aps,pra,showpacs,twocolumn]{revtex4}
\usepackage{epsfig}
\begin{document}
\title{Application of quantum algorithms to the study of permutations and
group automorphisms}
\author{Marianna Bonanome$^{1}$}
\author{Mark Hillery$^{1,2}$}
\author{Vladim\' \i r Bu\v{z}ek$^{3}$}
\affiliation{$^{1}$Department of Mathematics, Graduate Center
of the City University of New York, 365 Fifth Avenue, New York, NY 10016}
\affiliation{$^{2}$Department of Physics, Hunter College of
the City University of New York, 695 Park Avenue, New York, NY 10021}
\affiliation{$^{3}$Research Center for Quantum Information,Slovak Academy of
Sciences, D\'{u}bravsk\'{a} cesta 9, 845 11 Bratislava, Slovakia}

\begin{abstract}
We discuss three applications of efficient quantum algorithms to determining properties
of permutations and group automorphisms.  The first uses the
Bernstein-Vazirani algorithm to determine an unknown homomorphism from
$Z_{p-1}^{m}$ to $Aut(Z_{p})$ where $p$ is prime.
The remaining two make use of modifications of the Grover search algorithm.
The first finds the fixed point of a permutation or an automorphism
(assuming it has only one besides the identity).  It can be generalized to
find cycles of a specified size for permutations or orbits of a specified
size for automorphisms.  The second finds which of a set of permutations or
automorphisms maps one particular element of a set or group onto another.
This has relevance to the conjugacy problem for groups. We show how two of these algorithms can
be implemented via programmable quantum processors. This approach open new perspectives in quantum information processing
when both the data and the programs are represented by states of quantum registers. In particular, quantum programs that specify
control over data can be treated using methods of quantum information theory.
\end{abstract}
\maketitle

\section{Introduction}
Group theory has proven to be a useful arena in which to explore the application of quantum
algorithms, that is implementation of fundamental concepts of quantum theory for design of efficient
algorithms. Perhaps the most prominent example of this is the algorithm for the hidden subgroup
problem \cite{boneh,hoyer,mosca,jozsa,lomont}.
In this problem, we are given a black box that evaluates a function, $f$.  The function
maps a finite group $G$ to some finite set $X$. The function $f$ separates \cite{separate} cosets \cite{coset} of some subgroup $K$ of $G$.
That is, $f$ is constant on the left cosets of some subgroup,
$K$ of $G$ and distinct for different cosets.  The object is to use the black box
(the function f) to determine a generating set for the subgroup $K$.
This problem can be solved efficiently for Abelian groups by using a quantum
algorithm. Specifically,  Shor's algorithm for factoring is one particular 
example of this quantum algorithm. A quest for efficient quantum algorithms 
for the hidden subgroup problem for arbitrary groups would result in 
efficient quantum
algorithms for problems such as the graph isomorphism problem and the 
shortest vector problem  in lattices.
Recently, progress has been made for certain non-Abelian groups, one example
being the dihedral group \cite{kuperberg,regev,bacon} .

\subsection{Quantum algorithms}
Here we wish to show that quantum algorithms can be applied to the study of
permutations and  group automorphisms.  If $S$ is a finite set, a permutation,
$\sigma$ is a bijection mapping $S$ into itself.  A group automorphism is
a permutation on a group that satisfies certain conditions. In particular,
an automorphism, $\alpha :G\rightarrow G$ of a group, $G$, is a bijection
of the group back into itself so that the group operation is preserved, i.e.
if $g_{1},g_{2}\in G$, then $\alpha (g_{1}g_{2})=\alpha (g_{1})\alpha
(g_{2})$.  The set of automorphisms of a group, $G$, is itself a group,
denoted by $Aut(G)$, with the group operation being function composition.  If
the order of the group, $G$, is $N$, then the automorphism group of $G$ is a
subgroup of the symmetric group on $N$ objects, that is the group formed by
permutations of $N$ objects.

{\bf (1)} Our first algorithm solves the following problem by making use of a variant
of the Bernstein-Vazirani algorithm \cite{bernstein,cleve}.
Consider the cyclic group $Z_{p}$,
where $p$ is prime \cite{cyclic}.  The automorphism group of $Z_{p}$ is isomorphic to
$Z_{p-1}$ \cite{rotman}.  We consider a function $f : Z_{p-1}^{m}
\rightarrow Aut(Z_{p})$.  In particular, we have that
\begin{equation}
\label{fzmtoaut}
f: (n_{1}, n_{2}, \ldots n_{m})\rightarrow \alpha_{j_{m}}^{n_{m}}
\alpha_{j_{m-1}}^{n_{m-1}} \ldots \alpha_{j_{1}}^{n_{1}} ,
\end{equation}
where $n_{k}\in Z_{p-1}$ and $\alpha_{j_{k}}\in Aut (Z_{p})$.  We have a
circuit whose inputs are an element of $Z_{p-1}^{m}$, $\overline{n}
=(n_{1}, n_{2}, \ldots n_{m})$, and
an element, $l\in Z_{p}$, and whose output is an element of $Z_{p}$ given
by $\alpha (l)$, where $\alpha =f(\overline{n})$.  Our task is to determine
the automorphisms $\{\alpha_{j_{k}} |k=1,2,\ldots m\}$ with as few uses of
the circuit as possible.

The other two algorithms are both variants of the Grover search algorithm
\cite{grover}.\newline
{\bf (2)} The first finds fixed points of permutations or automorphisms.  Suppose the
permutation $\sigma$ on $S$ has only one fixed point, that is, there is only
one element $s\in S$ such that $\sigma (s)=s$.  The object is to find the
fixed point evaluating the permutation as few times as possible.  This can
clearly be used to find fixed points of automorphsims as well.  By applying
the same procedure to $\sigma^{n}$ we can also find elements of $S$
satisfying $\sigma^{n}(s)=s$.  Permutations can be written as the product
of disjoint cycles.  By searching for points satisfying $\sigma^{n}(s)=s$,
we are effectively searching for cycles of length $n$.  Therefore, this
algorithm can be used to find cycles of a specified length in a permutation.
For an automorphism, this corresponds to finding an orbit of a particular
length.

{\bf (3)} The final algorithm searches among permutations to find one with a specific
property.  In particular, suppose we have a set of permutations $\sigma_{k}$
where $k=1,2,\ldots M$.  Let $s_{1}$ and $s_{2}$ be two specfied elements of
$S$, and suppose that only one of the permutations satifies the condition
$\sigma_{k}(s_{1})=s_{2}$.  We want to find which one making use of as
few function evaluations as possible.  The corresponding problem for
automorphisms is related to the conjugacy problem.  For a group $G$, $g_{1}$
and $g_{2}$ in $G$ are said to be conjugate if there is an element $g_{0}$
of $G$ such that $g_{1}=g_{0}g_{2}g_{0}^{-1}$.  The mapping $\alpha_{g_{0}}
(g)=g_{0}gg_{0}^{-1}$ is an automorphism (automorphisms of this type are
known as inner automorphisms).  The conjugacy problem is, given two
elements of a group, to determine whether they are conjugate to each other.
The algorithm we are proposing can be applied to this problem.

Two of the algorithms we suggest are implemented by programmable quantum processors.
A programmable quantum processor is a device that can perform several different functions on
a quantum input state, which we shall call the data.  The operation that is performed on the
data is determined by a program, which is itself a quantum state.  The advantage of such a
device is that it is not necessary to construct a new quantum circuit for each operation, but only
to change the program.

\subsection{Programmable quantum processors}
Programmable quantum processors have been the object of a number of recent studies.  Nielsen
and Chuang showed that the number of unitary operations such a device can deterministically
perform is limited by the dimension of of the program space \cite{nielsen}.  This led to the
consideration of probabilistic \cite{nielsen,vlasov,vidal1,vidal2,hillery1,hillery2} and approximate processors
\cite{vlasov,vidal1,vidal2,hillery3}.  A probabilistic processor succeeds only part of the time, but we know
when it does and when it does not.  An approximate processor performs a set of operations to
some specified level of approximation.  Probabilistic and approximate processors do not suffer
from the same limitations as deterministic ones do, and can perform larger sets of operations
for a given program space dimension.

Another class of programmable devices consists of programmable measurement devices.  These
perform a measurement on an input state, with the measurement being specified by a program
state.  The first programmable measurement device was proposed by Du\v{s}ek and Bu\v{z}ek \cite{dusek},
and since then a number of different types of these devices have been studied
\cite{fiurasek,soubusta,sasaki,dariano,bergou,hayashi}.

One question that can be raised is whether having quantum programs gives one any advantage
over simply having classical ones.  A classically programmable quantum device can simply
be thought of as one with a dial, with different settings of the dial leading to different quantum
operations being applied to the input quantum state.  There are several reasons that quantum
programs can be advantageous.  First, the program itself may be the result of a previous quantum
computation.  This allows programmable quantum processors to be chained together, with the
output of one serving as the program of the next.  A second reason is that the program may
be intrinsically quantum.  As an example of this, consider the programmable measurement
device discussed in \cite{bergou}.  There the program consists of two qubits, one in the state
$|\psi_{1}\rangle$ and the other in the state $|\psi_{2}\rangle$, both of which are unknown.
The data consists of a qubit that is guaranteed to be in either $|\psi_{1}\rangle$ or
$|\psi_{2}\rangle$, and our task is to determine which.  Here, the program, consisting of two
qubits, is intrinsically quantum; we are comparing our unknown qubit to a two-qubit string in
order to determine which one it matches.  Finally, a third reason is that if the programs are
quantum, we can apply quantum information processing methods to the programs as well
as the data.  This is what we shall study here with two examples.

\section{Modified Bernstein-Vazirani algorithm}
Before proceeding to the modified version of the Bernstein-Vazirani algorithm
that solves the problem discussed in the Introduction, let us
review what the original algorithm does.  One is given a black box that evaluates a Boolean function, $f(x)$,
whose argument, $x=x_{n-1}\ldots x_{1}x_{0}$, is an $n$-digit binary number,
and $x_{j}=0,1$ for $0\leq j \leq n-1$.  The function is of the form
\begin{equation}
f(x)=\sum_{j=0}^{n-1}x_{j}y_{j}+b =x\cdot y +b ,
\end{equation}
where $y$ is a fixed, and unknown, $n$ digit binary number, $b=0$ or $1$,
and all additions and multiplications are modulo $2$.  The objective  is to find
$y$.  Classically this requires $n+1$ function evaluations.  One first finds
$b$ by evaluating the function for $x=0$, and then one finds $y_{j}$ by
choosing $x$ to be the string with $x_{j}=1$ and all of the other digits equal
to $0$.  Quantum mechanically only one function evaluation is required.

In the version of the algorithm considered here, the function is a mapping
from $Z_{p-1}^{m}$ to $Aut(Z_{p})$ rather than one from $Z_{2}^{n}$
to $Z_{2}$.
As was stated in the introduction, the automorphism group of $Z_{p}$, where
$p$ is prime, is isomorphic to $Z_{p-1}$.  This can be seen as follows.  Each
of the automorphism is completely determined by its action on $1$, so let us
denote by $\alpha_{k}$ the automorphism that satisfies $\alpha_{k}(1)=k$.
Note that $\alpha_{k}(n) = nk\ {\rm mod}k$.  There are clearly $p-1$ such
automorphisms, and any of the $\alpha_{k}$ for which $k$ does not divide
$p-1$ is a generator of the group $Aut(Z_{p})$.

We now have the function $f : Z_{p-1}^{m}\rightarrow Aut(Z_{p})$ specified
by Eq.\ (\ref{fzmtoaut}).  This function can be implemented by a quantum
circuit (see Fig.\ 1).  It consists of gates that are controlled-unitary
gates.  The control line has inputs from $Z_{p-1}$.  The target line has
inputs taken from $Z_{p}$.  Elements of both of these groups are encoded
as vectors from orthonormal bases.  The action of the controlled-unitary
gate is given by ($n\in Z_{p-1}$ and $l\in Z_{p}$)
\begin{equation}
|n\rangle |l\rangle \rightarrow |n\rangle
|\alpha_{k}^{n}(l)\rangle ,
\end{equation}
for some $\alpha_{k}$.  We have $p-1$ different kinds of controlled-unitary
gates, one for each automorphism in $Aut(Z_{p})$.  The circuit is composed
of $m$ of these gates. Consequently, there are $m$ control lines, one for
each of the inputs from $Z_{p-1}^{m}$.  All of the unitary operators act
on a common target line.  The first gate corresponds to $\alpha_{j_{1}}$,
the second to $\alpha_{j_{2}}$, etc.  The inputs to the circuit are
$|\overline{n}\rangle = |n_{1}\rangle \ldots |n_{m}\rangle$ corresponding
to an element of $Z_{p-1}^{m}$ and $|l\rangle$ corresponding to
an element of $Z_{p}$.  We shall denote the span of the vectors
$|\overline{n}\rangle$ by $\mathcal{H}_{P}$, the program space, and the span
of the vectors $|l\rangle$ by $\mathcal{H}_{D}$, the data space.
The action of the circuit is
\begin{equation}
|\overline{n}\rangle |l\rangle \rightarrow |\overline{n}\rangle
|\alpha (l)\rangle ,
\end{equation}
where $\alpha =f(\overline{n})$.

\begin{figure}[ht]
\epsfig{file=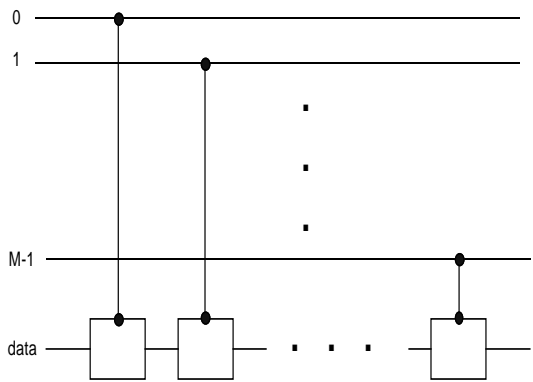,height=5cm,width=7cm}
\caption{The quantum circuit for the modified Berstein-Vazirani algorithm.
Each control line is a qudit, where $d=p-1$, that is connected to a gate that
implements a unitary operation on the data.  The data space is spanned by
the basis $\{ |l\rangle | l=1,2,\ldots p-1\}$, where each
basis state encodes a member of $Z_{p}$.  If the $k^{\rm th}$ control line
is in the state $|n\rangle$ the operation $U_{j_{k}}^{n}$ is performed, where
$U_{j_{k}}$ corresponds to the automorphism $\alpha_{j_{k}}$.  We have that
$U_{j_{k}}| l\rangle=|\alpha_{j_{k}}(l)\rangle$.}
\end{figure}

For each of the automorphisms, $\alpha_{k}$, define the vector in
$\mathcal{H}_{D}$
\begin{equation}
|u_{k}\rangle = \frac{1}{(p-1)^{1/2}}\sum_{n^{\prime}=0}^{p-2}
e^{-2\pi i n^{\prime}/(p-1)}|\alpha_{k}^{n^{\prime}}(1)\rangle .
\end{equation}
If $U_{k}$ is the operator whose action is given by $U_{k}|l\rangle
=|\alpha_{k}(l)\rangle$, then we have that
\begin{equation}
U_{_k}|u_{k}\rangle = e^{2\pi i/(p-1)}|u_{k}\rangle .
\end{equation}
We now choose one of the automorphisms, say $\alpha_{j_{0}}$, that is a
generator of $Aut(Z_{p})$, i.e. any member of $Aut(Z_{p})$ can be
expressed as a power of $\alpha_{j_{0}}$, and construct the vector
$|u_{j_{0}}\rangle$.  This vector is a simultaneous eigenvector of all
of the operators $U_{k}$.  In particular, if $\alpha_{j}=\alpha_{j_{0}}^{s}$,
then
\begin{equation}
U_{j}|u_{j_{0}}\rangle = e^{2\pi i s/(p-1)}|u_{j_{0}}\rangle .
\end{equation}
Therefore, there is a one-to-one correspondence between automorphisms and
eigenvalues of $|u_{j_{0}}\rangle$.

We take the input state to the quantum circuit to be
\begin{equation}
|\Psi_{in}\rangle =\left( \frac{1}{(p-1)^{m/2}}
\sum_{\overline{n}\in Z_{p-1}^{m}} |\overline{n}\rangle \right)\otimes
|u_{j_{0}}\rangle .
\end{equation}
The output state is then
\begin{equation}
|\Psi_{out}\rangle = \left( \frac{1}{(p-1)^{m/2}}
\sum_{\overline{n}\in Z_{p-1}^{m}}e^{2\pi i\overline{y}\cdot\overline{n}/(p-1)}
|\overline{n}\rangle \right) \otimes |u_{j_{0}}\rangle
\end{equation}
where $y_{k}$ is the unique integer between $0$ and $p-2$ such that
\begin{equation}
\alpha_{j_{k}}=\alpha_{j_{0}}^{y_{k}}
\end{equation}
We now note the following useful fact.  Defining the vectors
\begin{equation}
|v_{\overline{r}}\rangle =\frac{1}{(p-1)^{m/2}}
\sum_{\overline{n}\in Z_{p-1}^{m}}e^{2\pi i\overline{r}\cdot\overline{n}/(p-1)}
|\overline{n}\rangle ,
\end{equation}
where $\overline{r}\in Z_{p-1}^{m}$ and $\overline{r}\cdot\overline{n}=
r_{1}n_{1}+\ldots r_{m}n_{m}$, we have that
\begin{equation}
\langle v_{\overline{q}}|v_{\overline{r}}\rangle =
\delta_{\overline{q},\overline{r}} .
\end{equation}
Therefore, by measuring the output of the control lines in the basis
$|v_{\overline{r}}\rangle$, we can determine $\overline{y}$.  This
member of $Z_{p-1}^{m}$ completely specifies the function $f$, i.e. we
know which automorphism is performed by each control line in the circuit.

One use of the quantum circuit has allowed us to determine the unknown
function $f$.  Classically, $m$ uses would be required.  One would
always send $1$ into the target input, and $1$ into one of the control
inputs and $0$ into the others.  This would allow one to determine the
automorphism corresponding to the control input into which $1$ had been
sent.  Repeating this operation for each control line ($m$ times) would
allow one to determine $f$.

Finally, let us note that this algorithm solves a particular kind of hidden
subgroup problem, which is rather different from the usual one.  The mapping
$f$ is a homomorphism from $Z_{p-1}^{m}$ to $Aut(Z_{p})$, and so its
kernel is a subgroup of $Z_{p-1}^{m}$, and its image is a subgroup of
$Aut(Z_{p})$.  One can view the algorithm we have
just presented as solving the problem of finding either the kernel
or the image of the unknown homomorphism, $f$.  For example, suppose we
wish to find the kernel.  Once we have
found $\overline{y}$ with one use of the quantum circuit, the kernel is
given by the elements of $Z_{p-1}^{m}$ satisfying $\overline{y}\cdot
\overline{n}= 0\ {\rm mod} (p-1)$.

\section{Finding fixed points}
Let us now consider the following problem.  We are given a black box that
implements an unknown permutation on a finite set, $S$, i.e.
$\sigma : S \rightarrow S$.  We would like to find the fixed
points of the permutation, that is those elements $s\in S$ such that
$\sigma (s) =s$.  An algorithm that solves this problem could also be used
to find cycles.  Any permutation can be written as the product of disjoint
cycles.  If we know that a permutation, $\sigma$, has one cycle of $n$ and
has no cycles whose size divides $n$, then we can find the cycle by applying
the algorithm to $\sigma^{n}$, because the elements of the cycle are fixed
points of $\sigma^{n}$.  Group automorphisms are special kinds of
permutations, so they too can be decomposed into disjoint cylces.  In that
case the cycles are known as orbits.  Clearly an algorithm that solves the
problem of finding fixed points of permutations, would also be capable of
finding orbits of automorphisms.

Consider now a set $S$ and an permutation $\sigma$ of that set that has
only one fixed point, $s_{0}$. We shall show how we can use a modified Grover
algorithm to find the fixed point.  The elements of the set are encoded into an
orthonormal basis of a Hilbert space, $\mathcal{H}_{S}$ whose dimension is
equal to the number of elements in the set, $N$.  We shall be operating in the
space $\mathcal{H}_{a} \otimes \mathcal{H}_{b}$ where both
$\mathcal{H}_{a}$ and $\mathcal{H}_{b}$ are copies of $\mathcal{H}_{S}$ .
Define the state
\begin{equation}
|v\rangle_{ab} =\frac{1}{\sqrt{N}}\sum_{s\in S}|s\rangle_{a} \otimes
|s\rangle_{b}  ,
\end{equation}
and the operator $U_{\sigma}|s\rangle = |\sigma (s)\rangle$.  We begin the
iteration that
amplifies the amplitude of the state $|s_{0}\rangle$ by preparing the state
\begin{equation}
|\Psi_{in}\rangle_{ab} = (U_{\sigma}\otimes I_{b})|v\rangle_{ab} .
\end{equation}

We now need an operator that will perform the amplification of the
desired state.  This operator consists of two parts, one part flips the
sign of the target state and this is followed by what
Grover called the ``inversion about the mean''.  The operator that
performs the sign flip is $U_{id}=I_{ab}-2P_{id}$, where
\begin{equation}
P_{id} = \sum_{s\in S} (|s\rangle_{a}\langle s|)\otimes
(|s\rangle_{b}\langle s|) .
\end{equation}
Clearly, $U_{id}|s_{1}\rangle_{a}|s_{2}\rangle_{b}=|s_{1}\rangle_{a}
|s_{2}\rangle_{b}$ if $s_{1}\neq s_{2}$,
and $U_{id}|s\rangle_{a}|s\rangle_{b }= -|s\rangle_{a}|s\rangle_{b}$.  The
second operator is given by
\begin{equation}
U_{w}=(U_{\sigma}\otimes I_{b}) (I_{ab}-2|v\rangle_{ab}\langle v|)
(U_{\sigma}^{-1}\otimes I_{b} ) .
\end{equation}
Note that in order to implement this operator, we need black boxes that
implement both $U_{\sigma}$ and $U_{\sigma}^{-1}$.

We now apply $-U_{w}U_{id}$ to $|\Psi_{in}\rangle_{ab}$
$O(\sqrt{N})$ times. This has the effect of bringing the amplitude
of the state $|s_{0}\rangle_{a}|s_{0}\rangle_{b}$ close to one and
suppressing the amplitudes of all of the other states, so that if
the state is measured in the basis $\{ |s_{1}\rangle_{a}\otimes
|s_{2}\rangle_{b}| s_{1},s_{2}\in S\}$ the probability to find
$|s_{0}\rangle_{a}|s_{0}\rangle_{b}$ is close to one. We shall not
prove this statement here, because the argument is standard (see
Ref.~\cite{nielsenchuang}), and we shall present a more detailed
discussion of the Grover algorithm in the next section. It is
useful, however, to see the effect of one application of
$-U_{w}U_{id}$.  We have that
\begin{eqnarray}
-U_{w}U_{id}|\Psi_{in}\rangle_{ab} & = & \frac{1}{\sqrt{N}} \left[ \left( 3
- \frac{4}{N}\right)
|s_{0}\rangle_{a} |s_{0}\rangle_{b} \right.
\nonumber  \\
& +& \left. \left( 1-\frac{4}{N}\right) \sum_{s\neq s_{0}}
|\sigma (s)\rangle_{a} |s\rangle_{b} \right]  .
\end{eqnarray}
Note that the effect has been to increase the amplitude of
$|s_{0}\rangle_{a}|s_{0}\rangle_{b}$
and decrease the amplitude of all of the other states in the superposition.

If one does not know the number of fixed points, quantum counting can be used to find it
\cite{brassard, mosca2}.  This procedure uses phase estimation to find the eigenvalues
of the operator that implements the Grover iteration, and this allows one to determine the
number of solutions, which in this case is the number of fixed points.

\section{Determining whether two elements are automorphic images}
Let $X=\{ k|k=0,\ldots N-1 \}$ and suppose we have a set of permutations on $X$, $S=\{\sigma_{j}| j=1,\ldots M\}$.  In addition, if $x_{0},y_{0}\in X$ are specified elements of $X$, we shall assume
that only one of the permutations $\sigma_{j}$ satisfies the condition $\sigma_{j}(x_{0})=y_{0}$.
Our object is to find this permutation.

There are a number of problems for which the problem stated above forms a template.  We
are interested in the case in which the elements of the set $X$ are elements of a group, and
the permutations represent the actions of automorphisms, and $S$ is a subgroup of the automorphism group of the original group.  The object is then to find the automorphism for which
one specified group element is the automorphic image of another specified element.  The
problem can be modified so that we ask whether or not there is an automorphism in the set
$S$ so that one specified group element is the automorphic image of another.  The automorphisms
may or may not be inner.  An inner automorphism, $\alpha$, is of the
form $\alpha (g) = hgh^{-1}$, where $g$ is an arbitrary element of the
group $G$ and $h$ is fixed element of $G$.
If they are, we are trying to determine whether group elements are
conjugate to each other.  If $S$ is the subgroup of the automorphism group
consisting of the inner automorphisms, then we are trying to determine whether
two specified group elements are conjugate to each other, i.e.\ whether they
are in the same conjugacy class.  This is known as the conjugacy problem.

Another group theoretic problem which serves as a motivation to look at the
above search problem for permutations is the Whitehead problem.  In that
problem, one is trying to determine whether two elements of a free group
are automorphic images of each other \cite{lyndon}.  A generalized form
of the Whitehead problem considers $n$-tuples.  In particular, given two
$n$-tuples consising of elements of the free group, $(g_{1},g_{2},\ldots
g_{n})$ and $(h_{1},h_{2},\ldots h_{n})$, does there exist an automorphism
of the group, $\alpha$, such that $\alpha (g_{j})=h_{j}$ for $j=1,\ldots n$?
A further variant is to restrict the set of automorphisms one is considering.
In that case one would be interested in determining whether there is an
automorphism in the allowed set that maps one specified $n$-tuple to a
second specified one.

Classically, the only way to find the desired permutation is simply to try them one by one to
see which one gives the result $\sigma_{j}(x_{0})=y_{0}$.  This would typically take of the order
of $M$ steps.  What we want to show here is that by an application of a modified form of Grover's
algorithm, we can, on a quantum computer, find the desired permutation in a number of steps
that is of the order of $\sqrt{M}$.

We suppose that we have a quantum processor that causes the unitary operator $V$ to be applied
to the state vector $|\Psi\rangle$.  The Hilbert space is a product of two others, $\mathcal{H}=
\mathcal{H}_{S} \otimes \mathcal{H}_{X}$, where $\mathcal{H}_{S}$ is spanned by the
orthonormal basis $\{ |j\rangle_{S}| j=1,\ldots M\}$, and $\mathcal{H}_{X}$ is spanned by the
orthonormal basis $\{ |x\rangle_{X}| x=0,\ldots N-1\}$.  The operator $V$ has the following action
\begin{equation}
V|j\rangle_{S}|x\rangle_{X} = |j\rangle_{S} U_{j}|x\rangle_{X} ,
\end{equation}
where the operator $U_{j}$, acting on $\mathcal{H}_{X}$ implements the permutation
$\sigma_{j}$, i.e. $U_{j}|x\rangle_{X} =|\sigma_{j}(x)\rangle_{X}$.  What we see is that our
processor is a controlled-U gate, with the unitary operators that it performs corresponding to
the permutations in the set $S$.  As we shall see shortly, besides a processor that implements
$V$ we shall also need one that implements $V^{-1}$.

We start the system in the state $|\Psi_{in}\rangle =|w\rangle_{S} |x_{0}\rangle_{X}$, where
\begin{equation}
|w\rangle = \frac{1}{\sqrt{M}}\sum_{j=1}^{M}|j\rangle_{S} .
\end{equation}
We first apply $V$, and then apply the operator
\begin{eqnarray}
Q & =-&  V(I-2|w\rangle_{S}\langle w| \otimes |x_{0}\rangle_{X}\langle x_{0}|)
V^{-1} \nonumber \\
 & & (I-2I_{S}\otimes |y_{0}\rangle_{X}\langle y_{0}|) ,
\end{eqnarray}
a number of times $n$, where $n$ is yet to be determined, to the initial state.  At the end of the
iteration procedure our state is $Q^{n}V|\Psi_{in}\rangle$, and we then
measure the state in the basis $\{ |j\rangle_{S}| j=1,\ldots M\}$.  With a probability
close to one we will obtain the $j$ corresponding to the desired permutation.

In order to see how this works, let us find the state after one iteration. We can assume, without
loss of generatlity, that $j=1$ corresponds to the desired permutation,  We then have that
\begin{eqnarray}
Q|\Psi_{in}\rangle & = &V  \frac{1}{\sqrt{M}}\left[ \left( 3 -\frac{4}{M} \right) |1\rangle_{S} \right.
\nonumber \\
 & & \left. + \sum_{j=2}^{M}\left( 1- \frac{4}{M} \right) |j\rangle_{S} \right] \otimes |x_{0}\rangle_{X} .
\end{eqnarray}
We note that what has happened is that the probability of $j=1$ has increased, and the
probabilities of all of the other values of $j$ has decreased.  This is exactly what happens in
the standard Grover algorithm.  Further applications of $Q$ will increase the probability of
$j=1$ further, as long as we do not do it too many times.

For a more detailed analysis, we note, again as with the standard Grover algorithm, that all of the
action takes place in a two-dimensional subspace, and that $Q$ is simply a rotation in that
subspace.  In particular, the subspace is the span of the two vectors $|v_{1}\rangle =|1\rangle_{S}
|y_{0}\rangle_{X}$ and $|v_{2}\rangle =V |\Psi_{in}\rangle$.  We find that
\begin{eqnarray}
Q|v_{1}\rangle & = & |v_{1}\rangle -\frac{2}{\sqrt{M}}|v_{2}\rangle \nonumber \\
Q|v_{2}\rangle & = & \left( 1-\frac{4}{M}\right) |v_{2}\rangle
 + \frac{2}{\sqrt{M}}|v_{1}\rangle .
\end{eqnarray}
The vectors $|v_{1}\rangle$ and $|v_{2}\rangle$ are not orthogonal, so we define the vector
\begin{equation}
|v_{1}^{\perp}\rangle =\left(\frac{M}{M-1}\right)^{1/2} \left( |v_{2}\rangle -\frac{1}{\sqrt{M}}
|v_{1}\rangle \right) .
\end{equation}
In terms of the basis $\{ |v_{1}\rangle , |v_{1}^{\perp}\rangle \}$ the action of Q is given by
\begin{eqnarray}
Q|v_{1}\rangle & = & \left( 1-\frac{2}{M}\right) |v_{1}\rangle -\frac{2}{\sqrt{M}}\left( 1-\frac{1}{M}
\right)^{1/2}|v_{1}^{\perp}\rangle \, ;
\nonumber \\
\\
Q|v_{1}^{\perp}\rangle & = & \left( 1-\frac{2}{M}\right) |v_{1}^{\perp}\rangle
 +\frac{2}{\sqrt{M}}
\left( 1-\frac{1}{M}\right)^{1/2} |v_{1}\rangle  . \nonumber
\end{eqnarray}
Defining the angle $\alpha$ as
\begin{equation}
\sin \alpha = \frac{2}{\sqrt{M}} \left( 1-\frac{1}{M}\right)^{1/2} ,
\end{equation}
we see that for any vector of the form
\begin{equation}
|\psi\rangle =  \cos\theta |v_{1}\rangle +\sin\theta |v_{1}^{\perp}\rangle  ,
\end{equation}
we have that
\begin{equation}
Q|\psi\rangle = \cos (\theta -\alpha )|v_{1}\rangle + \sin (\theta -\alpha ) |v_{1}^{\perp}\rangle .
\end{equation}

The procedure we have outlined starts in the state $|v_{2}\rangle$, which is close to, but not quite,
orthogonal to $|v_{1}\rangle$.  We want to rotate the state so that it becomes close to $|v_{1}\rangle$, because it is the $\mathcal{H}_{S}$ part of $|v_{1}\rangle$ that contains the
information about which permutation has the desired property.  This means that we want to
iterate the process approximately $n=\pi /(2\alpha)\simeq  \pi \sqrt{M}/4$ times.  This is a
considerable improvement over the roughly $M$ steps we would have to perform
classically.

We can also easily extend this algorithm to search for $n$-tuples that are
automorphic images of each other as in the generalized Whithead problem.
Let us consider the case $n=2$.  In particular, we now want to find the
permutation, $\sigma_{j}$ such that $\sigma_{j}(x_{0})=y_{0}$ and $\sigma_{j}
(x_{1})=y_{1}$.  Our Hilbert space is now $\mathcal{H}=
\mathcal{H}_{S1}\otimes \mathcal{H}_{S2}\otimes \mathcal{H}_{X1} \otimes
\mathcal{H}_{X2}$, where $\mathcal{H}_{S1}$ and $\mathcal{H}_{S2}$ are
both copies of $\mathcal{H}_{S}$, and $\mathcal{H}_{X1}$ and
$\mathcal{H}_{X2}$ are copies of $\mathcal{H}_{X}$.  The operator $V\otimes V$
acts as
\begin{eqnarray}
(V\otimes V)|j_{1}\rangle_{S1}|j_{2}\rangle_{S2}|x\rangle_{X1}|x^{\prime}
\rangle_{X2} \nonumber \\
= |j_{1}\rangle_{S1}|j_{2}\rangle_{S2}U_{j_{1}}|x\rangle_{X1}
U_{j_{2}}|x^{\prime}\rangle_{X2} .
\end{eqnarray}
Now define the state
\begin{equation}
|W\rangle_{S1S2} = \frac{1}{\sqrt{M}}\sum_{j=1}^{M}|j\rangle_{S1}
|j\rangle_{S2} ,
\end{equation}
and the operators
\begin{eqnarray}
P_{X1X2}^{(x)}=|x_{0}\rangle_{X1}\langle x_{0}|\otimes |x_{1}\rangle_{X2}
\langle x_{1}|  \nonumber \\
P_{X1X2}^{(y)}=|y_{0}\rangle_{X1}\langle y_{0}|\otimes |y_{1}\rangle_{X2}
\langle y_{1}|   .
\end{eqnarray}
For our initial state we now choose $|\Psi_{in}\rangle = |W\rangle_{S1S2}
|x_{0}\rangle_{X1}|x_{1}\rangle_{X2}$.  We begin by applying $V\otimes V$
to this state, and then applying the operator
\begin{eqnarray}
Q & = & -(V\otimes V)(I-2|W\rangle_{S1S2}\langle W|\otimes P_{X1X2}^{(x)})
\nonumber \\
 & & (V^{-1}\otimes V^{-1})(I-2I_{S1S2}\otimes P_{X1X2}^{(y)}) ,
\end{eqnarray}
approximately $\pi\sqrt{M}/4$ times.  We then measure either the
$\mathcal{H}_{S1}$ or $\mathcal{H}_{S2}$ part of the state in the
computational basis to determine the permutation.

Note that in the algorithms being discussed in this section, the quantum
search is being applied to the programs that implement the permutations.
This would not be possible if those programs were classical.

\section{Conclusion}
We have presented three applications of efficient quantum algorithms to the study of group automorphisms.
In addition to finding properties of automorphisms, two of them can be applied to find properties
of permutations, i.e.\ finding fixed points and finding which permutation maps one particular
set element onto another.

This last task is accomplished by doing a Grover search on the programs of a programmable
quantum processor.  This shows that there are advantages to using quantum programs that
are themselves quantum states.  In particular, one can apply quantum information processing
to programs as well as data.  We expect that further work in this direction could prove useful.

{\bf Acknowledgement}\\
This work was supported in part by the European Union projects
CONQUEST and QAP,  by the Slovak Academy of Sciences via the
project CE-PI/2/2005, by the project APVT-99-012304.


\begin{thebibliography}{99}
\bibitem{boneh}R.\ Boneh and R.\ Lipton, in \emph{Lecture Notes in Computer Science} (Springer
Verlag, Berlin, 1995) p. 424.

\bibitem{hoyer} P.\ H\o yer, Phys.\ Rev.\ A {\bf 59}, 3280 (1999).

\bibitem{mosca} M.\ Mosca and A.\ Ekert, in \emph{Lecture Notes in Computer
 Science 1509}, edited by Colin Williams,  (Springer
Verlag, Berlin, 1999) p. 174, and quant-ph/9903071.

\bibitem{jozsa} R.\ Jozsa, Computing in Science \& Engineering {\bf 3}, 34 (2001), and quant-ph/0012084.

\bibitem{lomont} Ch.\ Lomont, quant-ph/0411037.

\bibitem{separate}
Given a group $G$, a subgroup $K\in  G$, and a set X, we say a function
$f : G \rightarrow X$ {\it separates} cosets of $K$ if for all $g_1, g_2 \in G, f(g_1) = f(g_2)$ if and only if $g_1 K = g_2 K$.

\bibitem{coset}
If $G$ is a group, $K$ a subgroup of $G$, and $g$ an element of $G$, then
$g K = \{ g k : k\in K \}$ is a left coset of $K$ in $G$, and
$K g = \{ k g : k \in K \}$ is a right coset of $K$ in $G$.
Only when $K$ is normal will the right and left cosets of $K$ coincide.

\bibitem{kuperberg} G.\ Kuperberg, SIAM J. Comp. {\bf 35}, 170 (2005), and quant-ph/0302112.

\bibitem{regev}O.\ Regev, quant-ph/0406151.

\bibitem{bacon}D. Bacon, A.\ M.\ Childs, and W.\ van Dam, Chicago Journal of
Theoretical Computer Science, no.\ 2, (2006), and quant-ph/0501044.

\bibitem{bernstein} E.\ Bernstein and U.\ Vazirani, Proc.\ 25th Annual ACM Symposium on
the Theory of Computing (ACM Press, New York, 1993) pages 11-20.

\bibitem{cleve} For a very nice discussion of the Berstein-Vazirani algorithm, and quantum
algorithms in general, see R.\ Cleve, A.\ Ekert, C.\ Macchiavello, and M.\ Mosca, Proc.\ R.\
Soc.\ Lond.\ A {\bf 454}, 339 (1998) and quant-ph/9708016.

\bibitem{cyclic}
A group $G$ is called {\it cyclic} if there exists an element $g$ (the generator) in $G$
such that, when written multiplicatively, every element of the group $G$ is a power of $g$.
Since any group generated by an element in a group is a subgroup of that group, showing that the only subgroup of a
group $G$ that contains $g$ is $G$ itself suffices to show that $G$ is cyclic.


\bibitem{rotman} Joseph J.\ Rotman, \emph{An Introduction to the Theory of
Groups} (Springer, New York, 1995).

\bibitem{grover} L.\ K.\ Grover, Phys.\ Rev.\ Lett.\ {\bf 79}, 325 (1997).

\bibitem{nielsen} M.\ A.\ Nielsen and I.\ L.\ Chuang, Phys.\ Rev.\ Lett.\ {\bf 79}, 321 (1997).

\bibitem{vlasov} A.\ Yu.\ Vlasov, e-print quant-ph/0103119.

\bibitem{vidal1} G.\ Vidal and J.\ I.\ Cirac, e-print quant-ph/0012067.

\bibitem{vidal2} G.\ Vidal, L.\ Masanes, and J.I.\ Cirac, Phys.\ Rev.\ Lett. {\bf 88}, 047905 (2002).

\bibitem{hillery1} M.\ Hillery, V.\ Bu\v{z}ek, and M.\ Ziman, Phys.\ Rev.\ A {\bf 65}, 022301 (2002).

\bibitem{hillery2} M.\ Hillery, M.\ Ziman, and V.\ Bu\v{z}ek, Phys.\ Rev.\ A {\bf 69}, 042311 (2004).

\bibitem{hillery3} M.\ Hillery, M.\ Ziman, and V.\ Bu\v{z}ek, Phys.\ Rev.\ A {\bf 73}, 022345 (2006).

\bibitem{dusek} M.\ Du\v{s}ek and V.\ Bu\v{z}ek, Phys.\ Rev.\ A {\bf 66}, 022112 (2002).

\bibitem{fiurasek} J.\ Fiura\v{s}ek, M.\ Du\v{s}ek, and R.\ Filip, Phys.\ Rev.\ Lett.\ {\bf 89}, 190401
(2002);  J.\ Fiura\v{s}ek and M.\ Du\v{s}ek, Phys.\ Rev.\ A {\bf 69}, 032302 (20040.

\bibitem{soubusta} J.\ Soubusta, A.\ \v{C}ernoch,  J.\ Fiura\v{s}ek, and M.\ Du\v{s}ek, Phys.
Rev.\ A {\bf 69}, 052231 (2004).

\bibitem{sasaki} M.\ Sasaki and A.\ Carlini, Phys.\ Rev.\ A {\bf 66}, 022303 (2002); M.\ Sasaki,
A.\ Carlini, and R.\ Jozsa, Phys.\ Rev.\ A {\bf 64}, 022317 (2001).

 \bibitem{dariano} G.\ M.\ D'Ariano and P.\ Perinotti, Phys.\ Rev.\ Lett.\ {\bf 94}, 090401 (2005).

\bibitem{bergou} J.\ Bergou and M.\ Hillery, Phys.\ Rev.\ Lett.\ {\bf 94}, 160501 (2005).

\bibitem{hayashi} A.\ Hayashi, M.\ Horibe, and T.\ Hashimoto, Phys.\ Rev. A {\bf 73}, 012328
(2006); Phys.\ Rev.\ A {\bf 72}, 052306 (2005).

\bibitem{nielsenchuang} M.\ A.\ Nielsen and I.\ L.\ Chuang, \emph{Quantum Computation and
Quantum Information} (Cambridge University Press, Cambridge, 2006) chapter 6.

\bibitem{brassard} G.\ Brassard, P.\ H\o yer, and A.\ Tapp, \emph{Lecture Notes in Computer Science 1443} (Springer Verlag, Berlin, 1999) p. 820,
and quant-ph/9805082.

\bibitem{mosca2} M.\ Mosca, in \emph{Proceedings of International Workshop on 
Randomized Algorithms, Workshop of Mathematical FOundations of Computer 
Science} edited by R.\ Freivalds, (Brno, Czech Republic 1998), available at
http://eccc.hpi-web.de/eccc-local/ECCC-LectureNotes/randalg/index.html.


\bibitem{lyndon} Roger C.\ Lyndon and Paul E.\ Schupp, \emph{Combinatorial
Group Theory} (Springer-Verlag, Berlin, 2001).

\end{thebibliography}
\end{document}